\documentclass[final,1p,times]{elsarticle} 

\usepackage{graphicx}
\usepackage{amssymb} 
\usepackage{amsthm} 

\journal{Nuclear Physics A} 
\usepackage{anysize}
\marginsize{3.5cm}{3.5cm}{3.2cm}{3.2cm}
\begin{document}

\begin{frontmatter} 
\title{\Large{Centrality dependence of freeze-out parameters from the \\beam energy scan at STAR}}
\author{\large{Sabita Das$^{a,b}$ (for the STAR Collaboration)}}
\fntext[col1] {sabita@rcf.rhic.bnl.gov}
\address[1]{ Brookhaven National Laboratory, Upton, NY, 11973-5000, USA}
\address[2]{ Institute of Physics, Bhubaneswar, 751005, INDIA}

\begin{abstract} 
  The STAR experiment at RHIC has a unique capability of measuring
identified hadrons over a wide range of pseudorapidity ($\eta$),
transverse momentum ($p_{T}$), and azimuthal angle ($\phi$)
acceptance. The data collected ($\sqrt{s_{NN}}$ = 7.7, 11.5, and 39 GeV)
 in its beam energy scan (BES) program provide a chance to investigate
 the final hadronic state freeze-out conditions of ultrarelativistic Au+Au
 collisions. The particle ratios are used to compare to a statistical
 model calculation using both grand canonical and strangeness
 canonical ensembles to extract the chemical freeze-out
 parameters. The $p_{T}$ distributions are fitted to calculations
 using a blast-wave model to obtain the kinetic freeze-out
 parameters. We discuss the centrality dependence of the
extracted chemical and kinetic freeze-out parameters at these lower energies.

\end{abstract} 
\end{frontmatter} 

\section{{Introduction}}
One of the early goals of heavy-ion collisions at Relativistic Heavy Ion
Collider (RHIC) is to establish the existence of a new state of matter
which is called quark-gluon plasma (QGP) [1]. According to lattice quantum
chromodynamics calculations (QCD), the theory of strong interactions,
QGP occurs when a sufficiently high temperature and high energy density
($\approx 1$ GeV/fm$^{3}$)
is reached. The partonic system is transformed into hadronic
matter as the system cools to lower temperatures. The phase diagram of
QCD is in general 
characterized by two quantities, the temperature (T) and the baryon
chemical potential ($ \mu_{B}$) or the (net) baryon
density ($n_{B}$). The phase diagram should contain information about
the phase boundary that separates the QGP and hadronic phases [2]. Lattice QCD finds a rapid,
but smooth crossover transition from hadron gas to QGP at vanishing
baryon chemical potential and large temperature T, while various models predict
a strong, first-order phase transition at large $ \mu_{B}$. If this is the
case, then there should be a critical point at intermediate values in the (T, $ \mu_{B}$ ) plane 
where the transition changes from a smooth crossover to a first order
[3]. The BES program at RHIC is carried out using several center-of-mass
energies of colliding nuclei, to explore the above
aspects of this QCD phase diagram.\\

The constituents of the hot and dense medium produced during a heavy-ion collision
interact with each other by inelastic and elastic collisions and it evolves into a state of free particles. This process of
hadron decoupling is called freeze-out. Two kinds of freeze-out are
found: chemical freeze-out (T$_{ch}$) when inelastic
collisions cease and the particle yields
become fixed; thermal (kinetic) freeze-out (T$_{kin}$) when
elastic collisions cease and particle transverse momenta ($p_{T}$)
spectra get fixed.\\
We present a study of the centrality dependence of hadronic freeze-out
parameters in Au+Au collisions at mid-rapidity for $\sqrt {s_{NN} }= $
7.7, 11.5, and 39
GeV measured by the STAR experiment. To extract 
chemical freeze-out parameters we use
a statistical thermal model (THERMUS) [5] where we fit experimental
particle ratios using a grand canonical ensemble (GCE) approach with the inclusion of
a strangeness saturation factor ($\gamma_{S}$) and also with
strangeness canonical ensemble (SCE) where strangeness quantum number
is conserved exactly. In this study we have used mid-rapidity particle ratios that
include the pions ($\pi^{+}$, $\pi^{-}$), kaons ($\emph{K}^{+}$,
$\emph{K}^{-}$), protons ($\emph{p}$, $\bar{\emph{p}}$),
${K^{0}}_{S}$, Lambdas ($\Lambda$, $\bar{\Lambda}$) and Cascades
($\Xi^{-}$, $\bar{\Xi}^{+}$) [6, 7]. The chemical freeze-out
parameters extracted are T$_{ch}$, $ \mu_{B}$, $ \mu_{S}$ and
$\gamma_{S}$ . The kinetic freeze-out parameters are determined from the blast-wave
model (BW) [8] fits to the $p_{T}$ spectra of $\pi$,
$\emph{K}$ and $\emph{p}$ [6]. The main kinetic freeze-out parameters
extracted are T$_{kin}$ and average flow velocity ($\langle \beta \rangle$). 
\section{Results}
\subsection{Kinetic freeze-out}
\begin{figure}
\begin{tabular}{cc}
\hspace{0.3cm}
\includegraphics[width = 2.5in,height=2.2in]{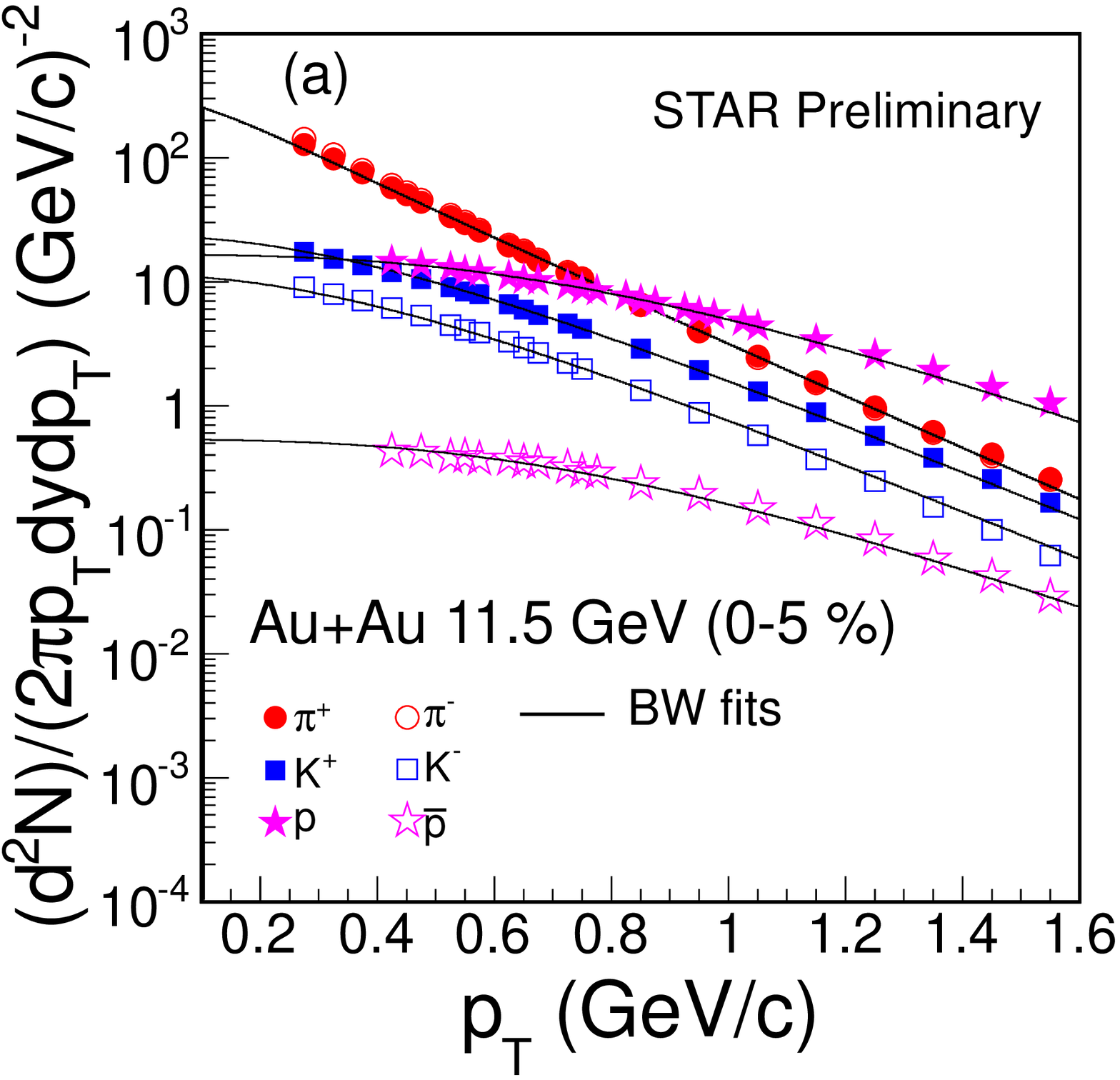} &
\includegraphics[width = 2.5in,height=2.2in]{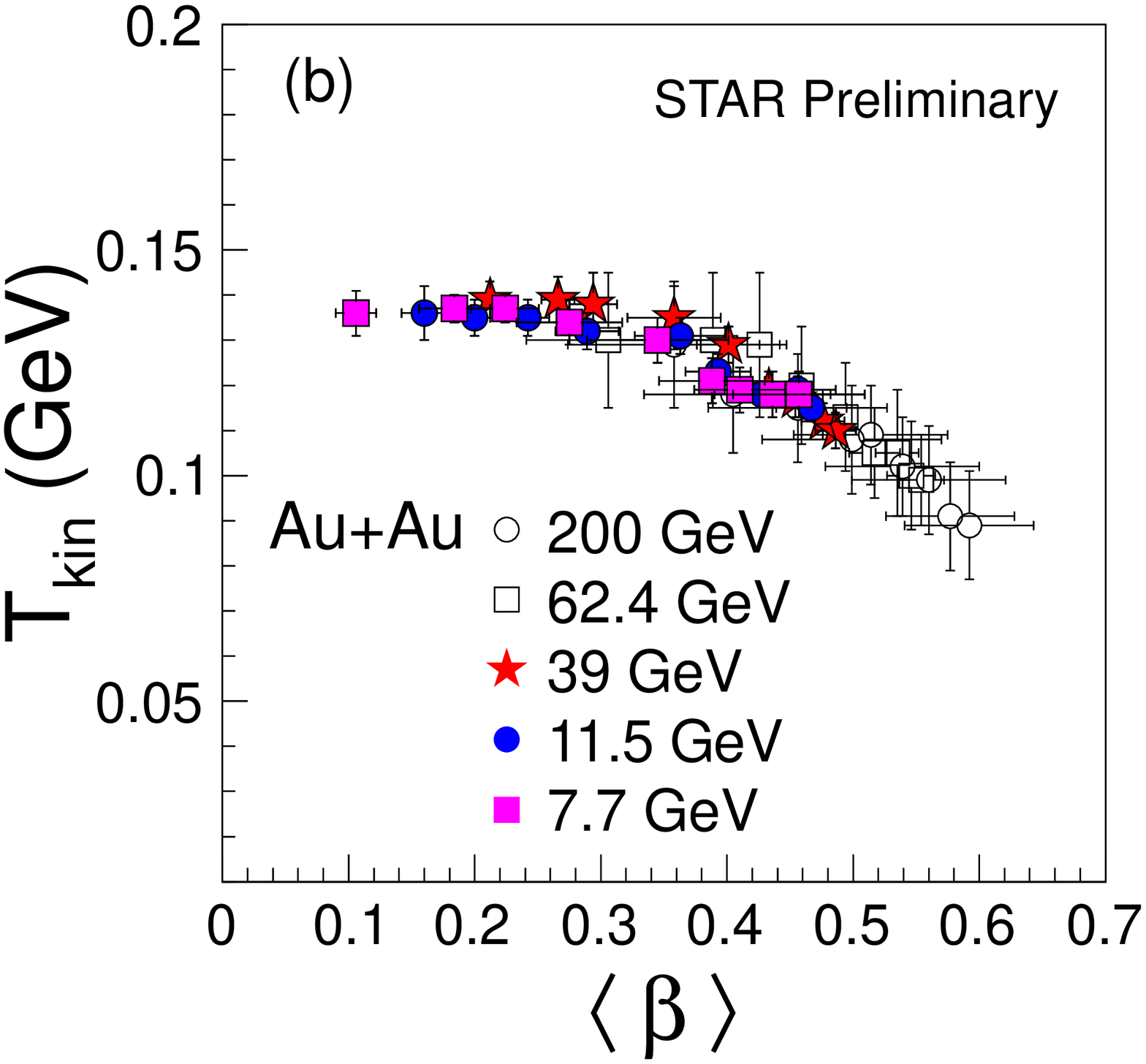}
\end{tabular}
\caption{(a) Simultaneous blast-wave fit of charged pion, kaon and
  proton for 0-5\% centrality in Au+Au collisions at $\sqrt{s_{NN}}$=
  $11.5$ GeV. (b) Variation of T$_{kin}$ with $\langle \beta \rangle$ for different centralities at 
$\sqrt{s_{NN} }$= 7.7, 11.5, 39, 62.4, and 200 GeV. The 62.4 and 200 GeV results are taken
from Ref.[4]. The errors shown here are the statistical and
systematic errors added in quadrature.}
\label{fig}
\end{figure}
At the kinetic freeze-out, elastic collisions among the particles stop and
the spectral shape of the particles get fixed. Kinetic freeze-out parameters
are obtained using blast-wave model by doing the simultaneous fits of
$\pi$, $\emph{K}$, and $\emph{p}$  transverse momentum spectra. The BW model describes the
spectral shapes assuming a locally thermalized source with a common transverse flow velocity field. It has been
successfully used to describe $p_{T}$ spectra with three parameters -
T$_{kin}$, $\langle\beta\rangle$, and the exponent in the flow
velocity profile $\emph {n}$ at 62.4 and 200 GeV [4, 9]. Figure 1(a) shows the simultaneous blast-wave fit
of $\pi$, $\emph{K}$, $\emph{p}$  and the corresponding antiparticles
for $0-5\%$ centrality in Au+Au collisions at $\sqrt{s_{NN}}$ =
  $11.5$ GeV. The variation of T$_{kin}$  as a
  function of $\langle \beta \rangle$ at $\sqrt{s_{NN} }$= 7.7, 11.5,
  39, 62.4 and 200 GeV is shown in Fig 1(b). The 62.4 and 200 GeV
  results are taken
from the Ref.[4]. The T$_{kin}$  decreases from peripheral to central
collisions. It also decreases with increasing collision energy. The
$\langle \beta \rangle$ increases with increase of energy as well as collision
centrality. So, higher value of  T$_{kin}$ corresponds to lower
value of $\langle \beta \rangle$ and vice-versa. The errors shown are the quadratic sum
of statistical and systematic errors.
\begin{figure}
\centering
\begin{tabular}{c}
\includegraphics[width = 2.6in,height=2.1in]{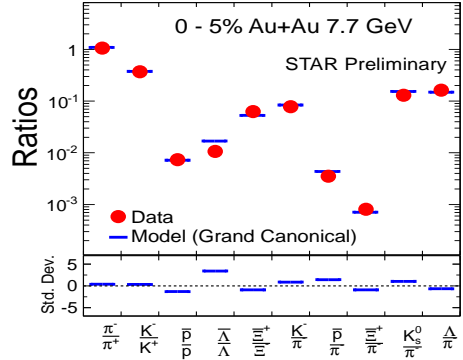} 
\end{tabular}
\caption{Statistical thermal model [5] fits to experimental
  mid-rapidity particle ratios for 0-5\% centrality in Au+Au 7.7 GeV energy.}
\label{fig}
\end{figure}
\subsection{Chemical freeze-out}
At the chemical freeze-out, inelastic collisions among the particles stop and
hadron yields get fixed. A statistical thermal model (THERMUS) was used to fit
mid-rapidity particle ratios including yields of  
$\pi$, $\emph{K}$, $\emph{p}$, ${K^{0}}_{S}$, $\Lambda$ and $\Xi$ measued in Au+Au collisions at
$\sqrt{s_{NN} }$= 7.7, 11.5, and 
  39 GeV.  Although the
  particle ratios are obtained at y=0, the measurements of yields for $\pi$, $\emph{K}$, $\emph{p}$ are for $|y| < 0.1 $ and those
  for ${K^{0}}_{S}$, $\Lambda$ , $\Xi$ are for $|y| < 0.5 $. The errors on particle ratios including yields of $\pi$,
  $\emph{K}$, $\emph{p}$, ${K^{0}}_{S}$, $\Lambda$, and $\Xi$, are the quadratic sum of statistical
and systematic uncertainties. Pion yields
have been corrected for feed-down from
${K^{0}}_{S}$ weak decays. Proton yields have not been corrected for
feed-down contributions. The $\Lambda$ yields have been
corrected for the feed-down contributions from $\Xi$ and $\Xi^{0}$ weak
decays [7]. In the framework of this model, the particle yield ratios
  can be described by a set of parameters such as T$_{ch}$, $\mu_{B}$, $\mu_{S}$
  and $\gamma_{S}$. The errors on
freeze-out parameters are obtained from THERMUS
model. Considering
  grand canonical formulation of this model we have studied the centrality
  and energy dependence of the freeze-out parameters. 

\begin{figure}
\begin{tabular}{ccc}
\hspace{-0.3cm}
\includegraphics[width = 1.97in,height=1.92in]{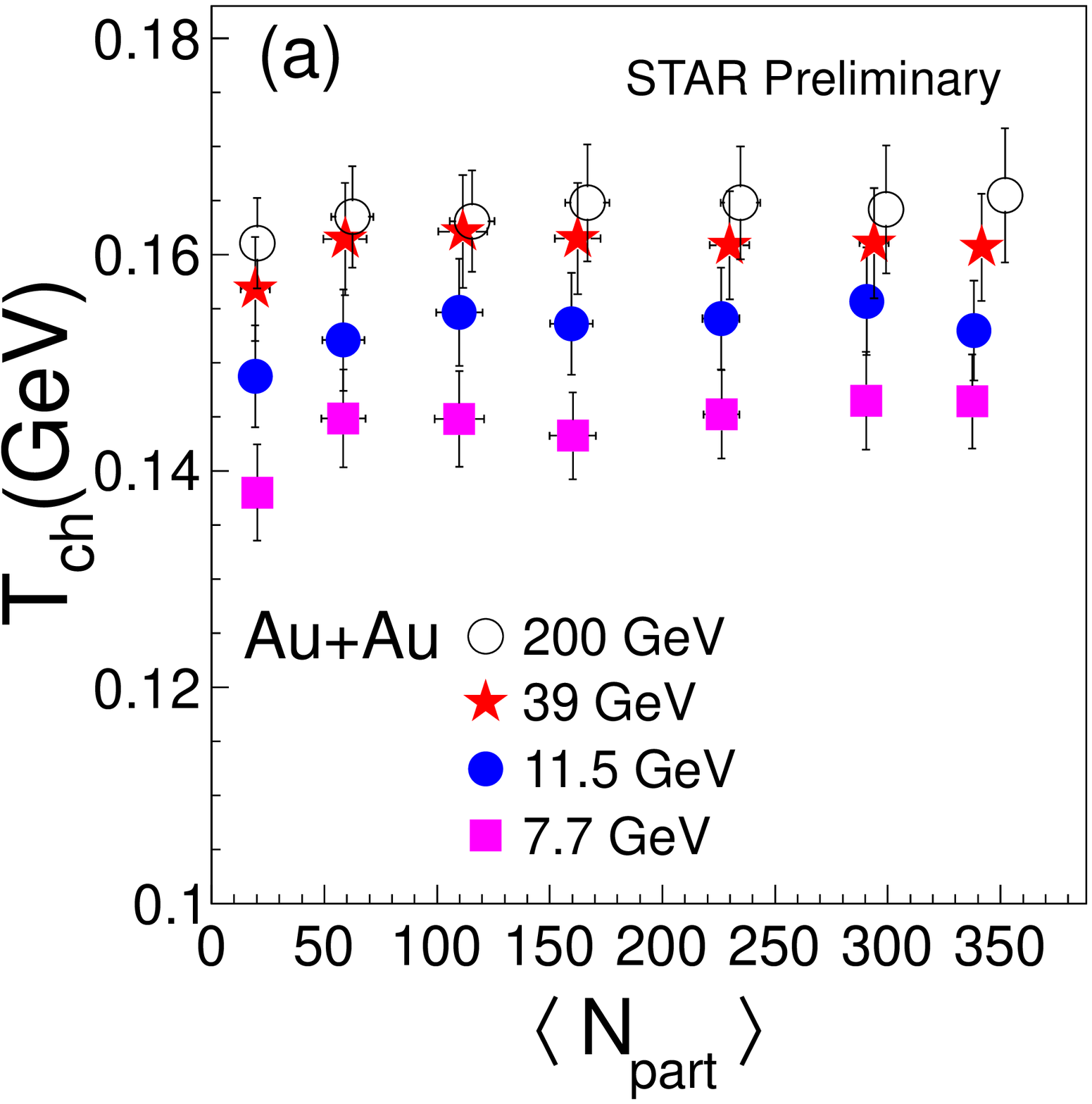} &
\hspace{-0.75cm}
\includegraphics[width = 1.97in,height=1.85in]{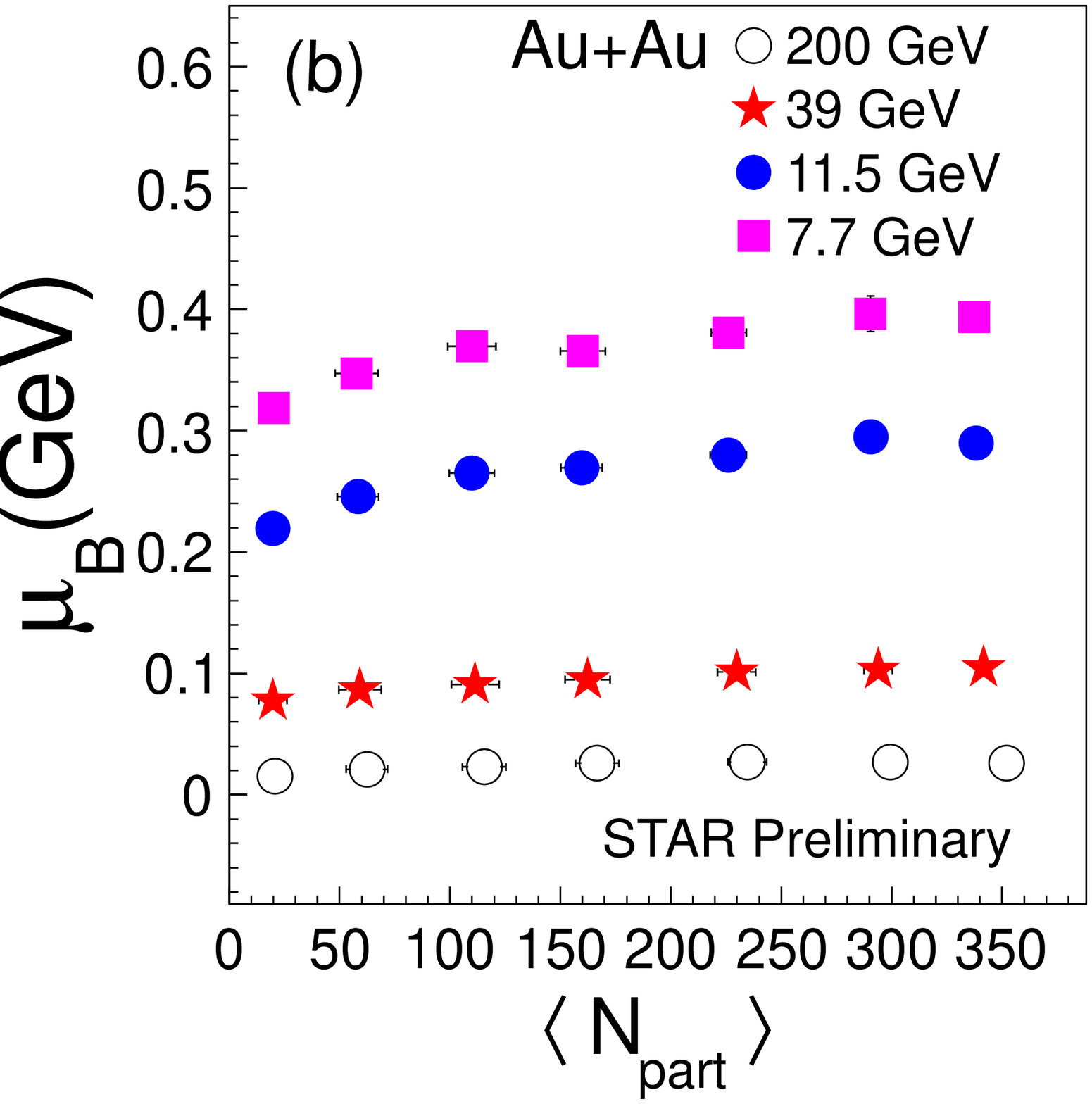}&
\hspace{-0.75cm}
\includegraphics[width = 1.95in,height=1.9in]{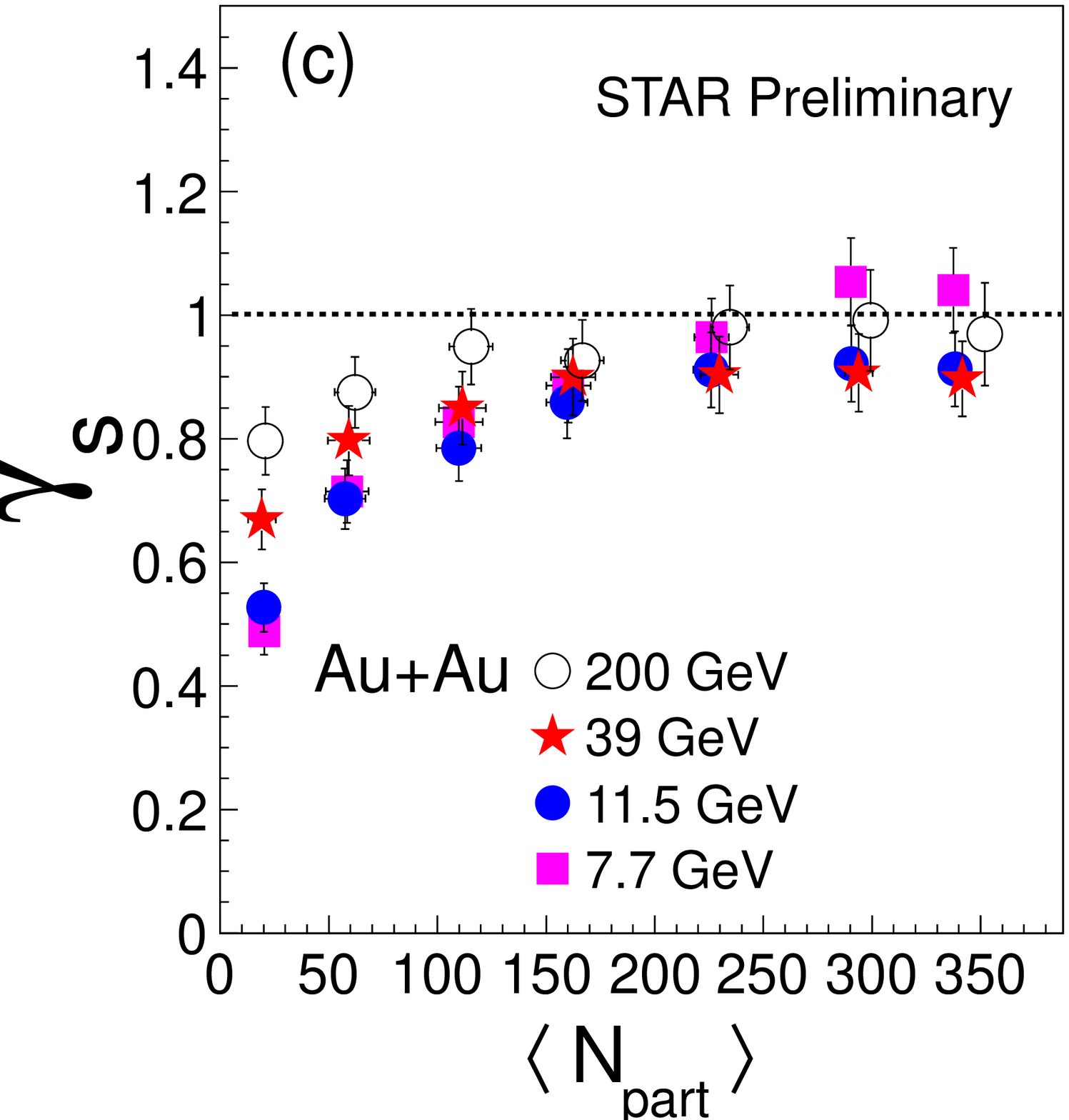}
\end{tabular}
\caption{Results from a statistical thermal model fit [5] for Au+Au
  7.7,11.5, and 39 GeV. Chemical freeze-out
temperature, baryon chemical potential, and strangeness saturation factor are
shown as a function of $N_{part}$. The 200 GeV results are taken from
Ref. [10].}
\label{fig}
\end{figure}
\vspace{0.07cm}
Figure 2 shows
  the statistical model fits to experimental particle ratios for
  $0-5\%$ centrality in Au+Au collisions at $\sqrt{s_{NN}}$ = 7.7
  GeV. The data and model matches very well, except for
  $\bar{\Lambda}/\Lambda$. Figure 3 (a) shows the T$_{ch}$ increases
  with increase of collision energy. Figure
  3(b) shows $\mu_{B}$ decreases with increasing collision
  energy. From peripheral to central collisions, $\mu_{B}$ 
  increases at lower energies. We observe a centrality dependence of chemical freeze-out
  curve (T$_{ch}$ vs. $\mu_{B}$) at BES energies which was not observed at
  higher energies like Au+Au 200 GeV [10]. Figure 3 (c) shows that the
  $\gamma_{S}$ increases from peripheral to central collisions for all
  energies.\\
 In contrast to GCE where all quantum numbers are conserved on an
 average, THERMUS model also allows for a strangeness canonical
 ensemble where only the strangeness quantum number is required to be
 conserved exactly where as baryon and charge quantum numbers are conserved
 on an average. Figure 4
 (a) and (b) indicate that in peripheral collisions, T$_{ch}$ and $\mu_{B}$ 
  follow a different behavior in GCE and SCE at  $\sqrt{s_{NN}}$ = 7.7 GeV. We observe a higher
 $\frac{\chi^{2}}{ndf}$ in SCE in comparison to GCE at pheripheral
 collisions. Further systematic investigations are ongoing towards a more quantitative analysis for all the BES energies.  
\begin{figure}
\begin{tabular}{ccc}
\hspace{-0.3cm}
\includegraphics[width =
1.97in,height=1.9in]{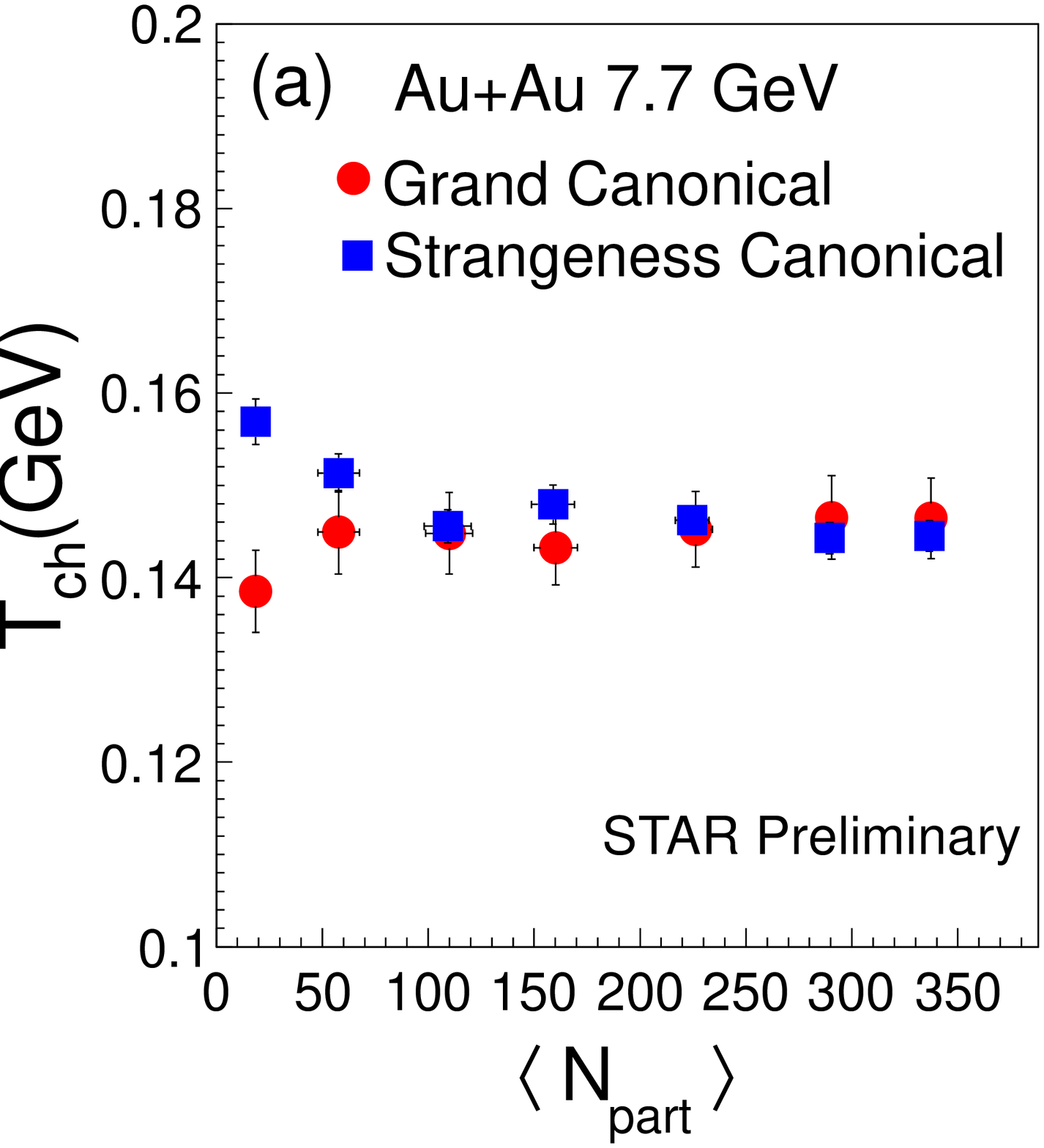}
&
\hspace{-0.75cm}
\includegraphics[width =
1.97in,height=1.9in]{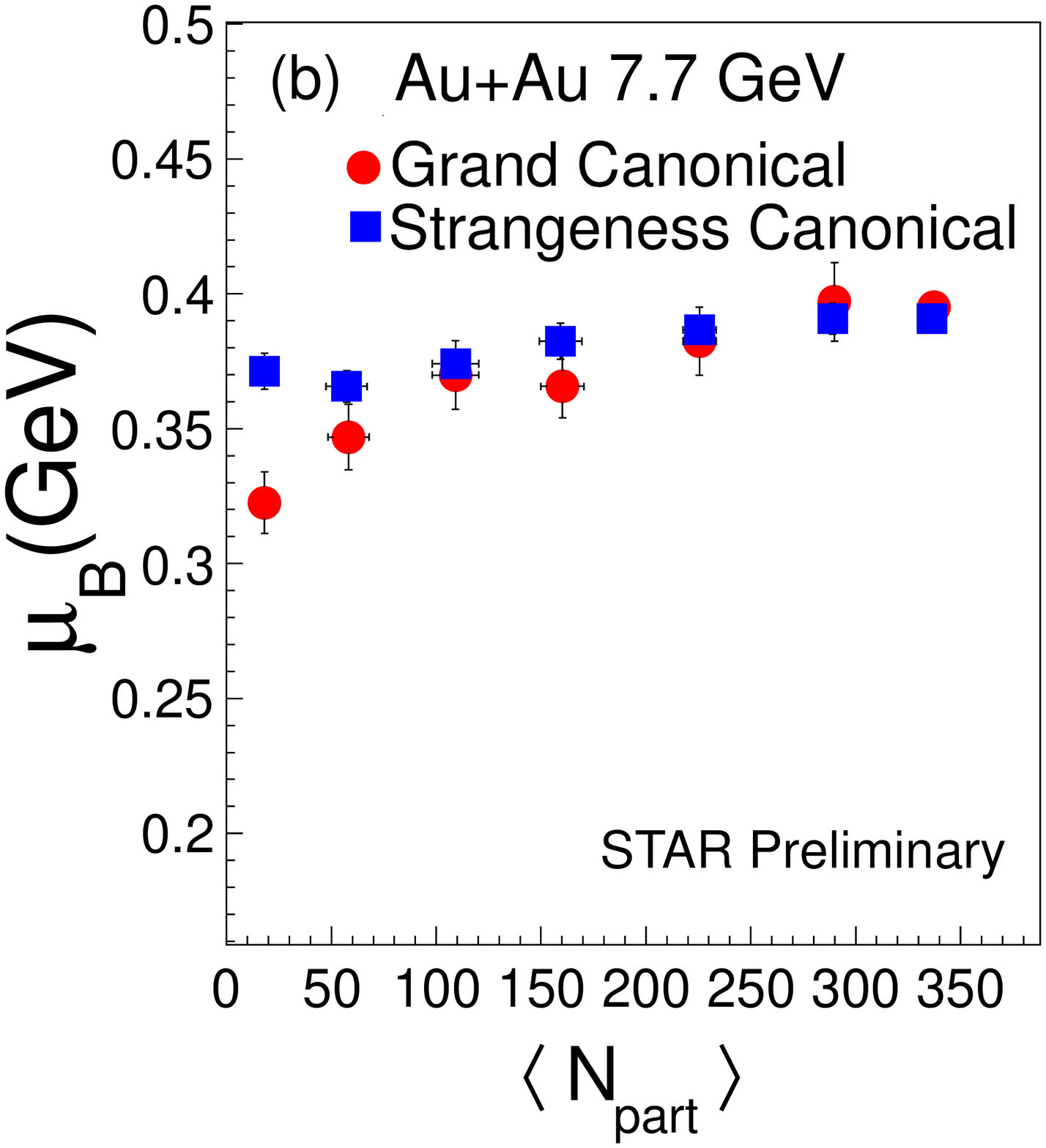}&
\hspace{-0.75cm}
\includegraphics[width = 1.97in,height=1.9in]{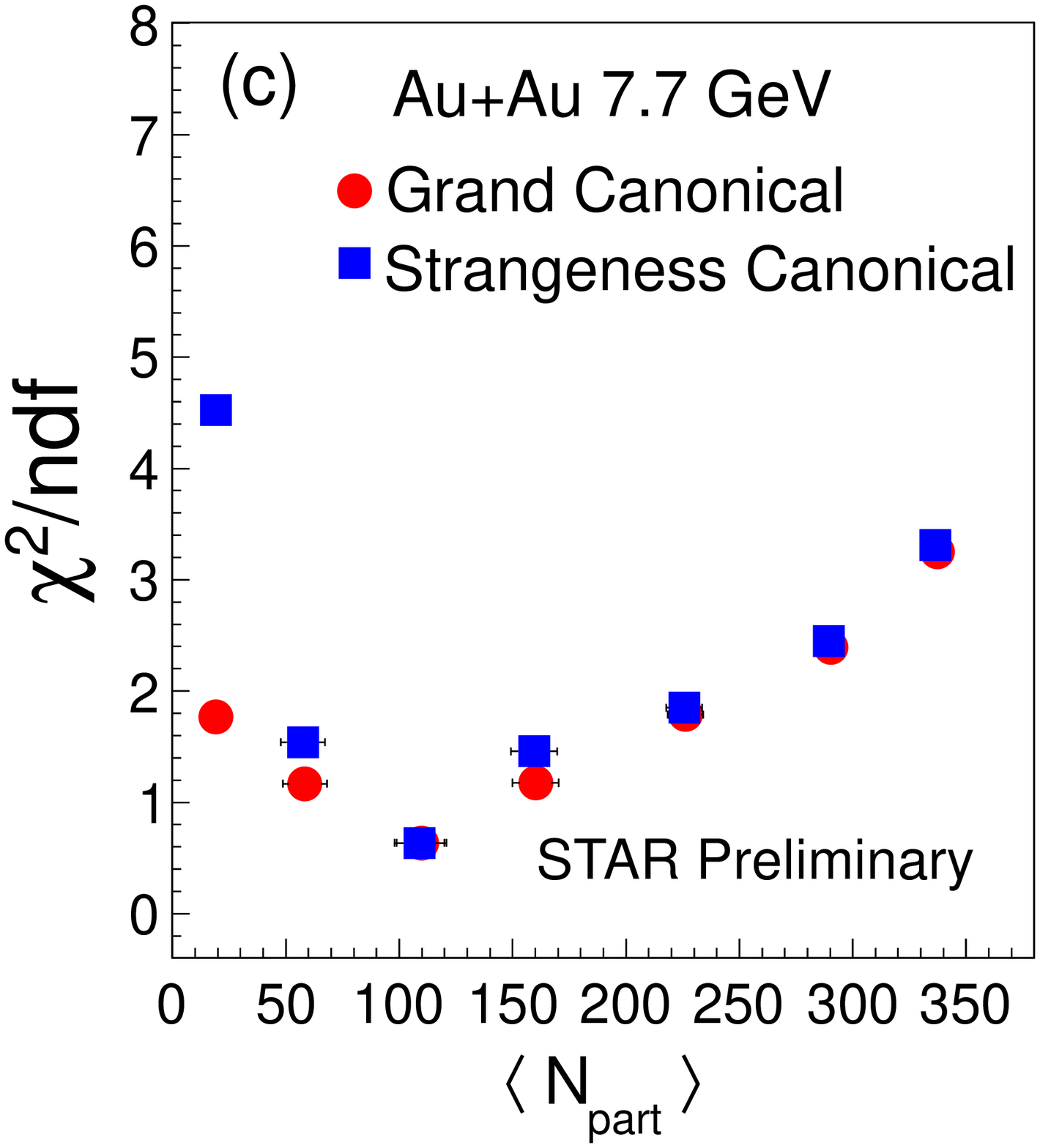}
\end{tabular}
\caption{Results from a statistical thermal model fit [5] for Au+Au
  7.7 GeV in GCE and SCE. Chemical freeze-out
temperature, baryon chemical potential and chisquare per degrees of freedom are
shown as a function of $N_{part}$.}
\label{fig}
\end{figure}
\section{Summary}
The new measurement at BES energies 7.7, 11.5, and 39 GeV at RHIC
extends the 
$\mu_{B}$ range from 20  to 400 MeV in the QCD phase diagram. 
Kinetic freeze-out parameters are obtained
using the measured particle spectra and a BW model. For all
the beam energies studied, the central collisions are characterized
by a lower T$_{kin}$ and larger $\langle \beta \rangle$ while the peripheral
collisions are found to have a higher  T$_{kin}$ and smaller $\langle \beta \rangle$. Chemical
freeze-out parameters are obtained using the measured particle ratios
and a THERMUS model. We have observed a centrality dependence of the chemical freeze-out parameters at
the lower energies. We have observed different behavior of chemical freeze-out
parameters (T$_{ch}$, $\mu_{B}$) for peripheral collisions in GCE and SCE.


\section*{References}

\begin{thebibliography}{00} 
\bibitem{Julius1} J. Adams et al. (STAR Collaboration), Nucl. Phys. A757, 102 (2005).
\bibitem{Julius2} P Braun-Munzinger et al. arXiv:1101.3167, 2011; B.Mohanty, Nucl. Phys. A 830, 899C (2009).
\bibitem{Julius3} S. Gupta et al., Science 332, 1525 (2011); E. S. Bowman and J. I. Kapusta, Phys. Rev. C 79, 015202 (2009).
\bibitem{Julius4} B. I. Abelev et al. (STAR Collaboration), Phys. Rev. C 79, 034909 (2009).
\bibitem{Julius5} J. Cleymans et al., Computer Physics Communications,
  180, 84 (2009).
\bibitem{Julius6} L. Kumar (STAR collaboration), arXiv:1201.4203
  (2012), J. Phys. G: Nucl. Part. Phys. 38, 124145 (2011).
\bibitem{Julius7} X. Zhu (STAR Collaboration), Acta Phys. Polon. B Proc. Supp. 5
(2012) 213-218.
\bibitem{Julius8} E. Schnedermann et al., Phys. Rev. C 48, 2462 (1993).
\bibitem{Julius9} B. I. Abelev et al. (STAR Collaboration), Phys. Rev. C 81, 024911 (2010).
\bibitem{Julius10} M. M. Aggarwal et al. (STAR Collaboration), Phys. Rev. C 83, 024901 (2011).
\end{thebibliography}
\end{document}